\documentclass[%
 reprint,
 amsmath,amssymb,
aps,
]{revtex4-2}
\usepackage{graphicx}
\usepackage{dcolumn}
\usepackage{bm}

\begin{document}

\preprint{APS/123-QED}

\title{Unified probability explanation for ghost imaging with thermal light}

\author{Wen-Kai Yu,$^{1,2,\dag,}$}
\email{yuwenkai@bit.edu.cn}
\affiliation{$^1$Center for Quantum Technology Research, School of Physics, Beijing Institute of Technology, Beijing 100081, China\\
$^2$Key Laboratory of Advanced Optoelectronic Quantum Architecture and Measurement of Ministry of Education, School of Physics, Beijing Institute of Technology, Beijing 100081, China\\
$^\dag$These authors contributed equally to this work\\}

\author{Jian Leng,$^{1,2,\dag}$}
\affiliation{$^1$Center for Quantum Technology Research, School of Physics, Beijing Institute of Technology, Beijing 100081, China\\
$^2$Key Laboratory of Advanced Optoelectronic Quantum Architecture and Measurement of Ministry of Education, School of Physics, Beijing Institute of Technology, Beijing 100081, China\\
$^\dag$These authors contributed equally to this work\\}



\date{\today}

\begin{abstract}
Ghost imaging (GI) is an intriguing imaging technology which achieves the object images through intensity correlation between reference patterns and bucket signal. Here, we propose a probability model to explain the imaging mechanism of this modality, by assuming that the reference patterns fulfill an arbitrary identical distribution and that the objects are of gray-scale. We have proven that the probability of the reconstructed pixel values in the pixel region of the same original gray value obeys a Gaussian distribution, no matter which functional form of the reference patterns is used in correlation calculation. Both simulation and experiments have demonstrated that the probability of recovered pixel values are highly consistent with their Gaussian theoretical distribution, while their variance explains the appearance of reconstruction noise. In addition, we have also extend this theory to other classic correlation functions, e.g., normalized GI and differential GI. The results have shown that there is a linear relationship between reconstruction means in specified pixel regions and original gray values, which might provide a unified explanation for GI with thermal light.
\end{abstract}

\maketitle


\section{\label{sec:level1}Introduction}
Ghost imaging (GI) retrieves the object information via intensity correlation between two separated but related light fields, i.e., object arm and reference arm. To our knowledge, the concept of GI was theoretically proposed by Belinskii and Klyshko in 1994 \cite{Belinskii1994}, and the first GI experiment was implemented in 1995 with quantum entangled photon pairs \cite{Pittman1995}, making GI be originally claimed as a quantum phenomenon. Later, it was shown that the classic thermal or pseudo-thermal light \cite{DaZhang2005,Ferri2005,Jun2005,Liu2014} could also be used to acquire the ghost images, arising a controversy on whether the entanglement is necessary for GI. One thing is certain: thermal or pseudo-thermal light has become a favorable source for GI, given its convenience in practical applications \cite{YuOC2016,Clemente2010,YuAO2019,Gong2015,YuSR2014}. If one uses a programmable spatial light modulator (SLM) or digital micromirror device (DMD) to replace the reference arm, then the pixelated array detector can be removed, which simplifies the imaging configuration. This technique was called computational GI \cite{Shapiro2008,Bromberg2009}. In recent years, GI has attracted more and more attention, and in order to improve its imaging quality and efficiency, many efforts have been made in its correlation reconstruction algorithms, such as background-removal GI \cite{Gatti2004}, high-order GI \cite{Chan2009}, differential GI (DGI) \cite{Ferri2010}, sequential-deviation GI \cite{LiOE2019}, super sub-Nyquist GI \cite{YuSensors2019}, etc.

However, the imaging mechanism of these algorithms lacks a universal unified interpretation. Inspired by Yang's theoretical interpretation work \cite{CaoPRA2018} which mainly focused on correspondence imaging and required a strong assumption, we build a new unified theoretical model where the speckle patterns can obey any identical distribution and the objects can be of gray-scale. We assume that the intensities of any two pixels in one speckle pattern as well as the ones on a certain pixel of different speckle patterns are independently and identically distributed, all following the same distribution. That is, any two pixel values of reference patterns in both space and time dimensions are independently and identically distributed, and each pixel value can be regarded as a stochastic variable. Then, the single-pixel (bucket) signal (also a random variable) can be treated as linear combinations of all these pixels. With these assumptions, we deduce an independence discrimination separated formula, which is applied to prove that the results of multi-functional intensity correlation forms are the linear transformation of the original gray values in terms of mean, thus the object information can be extracted. Both simulation and experiments have been performed to verify the correctness of this theoretical model, and to demonstrate that the recovered values via intensity correlation in the pixel region of the same original gray value will obey a Gaussian distribution.

\section{\label{sec:level2}Probability theory for ghost imaging}
The total pixel number of the object image can be denoted by $M$, and the gray value of one pixel is expressed by $d$. Then, we denote the gray value of the $m$th pixel as $d_m$, with a range covering from 0 to 1, where 0 stands for being completely opaque and 1 stands for being completely transparent. Assume that each reference speckle pattern also has $M$ pixels, and accordingly, the light intensity of the $m$th pixel is denoted by $I_m$. Suppose that the intensities of any two pixels in one reference pattern are independently and identically distributed, and those on a certain pixel of any two different reference patterns are also independently and identically distributed. They all follow an identical probability distribution $\mathcal{I}$, with a mean $E(\mathcal{I})$ and a variance $D(\mathcal{I})$. For the sake of generality, in this paper we will mainly discuss a functional form of $\mathcal{I}$ for reconstruction, i.e., $\mathcal{F}=f(\mathcal{I})$ \cite{Zhang2019}, instead of $\mathcal{I}$, where $f$ can be a power function, exponential function, logarithmic function, etc. Here, the function for the $m$th pixel is written as $F_m$, obeying a distribution $\mathcal{F}$.

After each reference pattern interacts with the object, the light intensity at the $m$th pixel of the spatial light field can be written as $\gamma d_mI_m$, where the factor $\gamma$ is a constant. The bucket signal can be acquired via
\begin{equation}
S=\gamma\sum_{m}^Md_mI_m.
\end{equation}
In the following, we need to divide $S$ into two parts, $\widetilde{S}$ and $\gamma d_nI_n$, i.e.,
\begin{equation}
S=\widetilde{S}+\gamma d_nI_n,
\end{equation}
where $\widetilde{S}=\gamma\sum_{m\ne n}^M d_mI_m$. Noticeably, $\widetilde{S}$ is independent of $I_n$. If $I_n$ is replaced with $F_n$, then $\widetilde{S}$ is also independent of $F_n$.

The second-order correlation function for the $n$th pixel can be written as
\begin{equation}
G^{(2)}_{n}=\langle S\cdot I_n\rangle=\frac{1}{T}\sum_t^T S_tI_{tn},
\end{equation}
where $T$ is the total number of reference patterns, the subscript $t$ stands for the $t$th measurement, and the subscript $tn$ denotes the $n$th pixel for the $t$th reference pattern. According to Liu's work \cite{Zhang2019}, $I_n$ in correlation function can be replaced with some functions $F_n$. Considering generality, we use $F_n=f(I_n)$ to represent an arbitrary function expect for constant function. With this assumption, the function $F$ can also be treated as a random variable, obeying a distribution $\mathcal{F}$. Then, we have
\begin{equation}
G^{(2)}_{n}=\langle S\cdot F_n\rangle=\frac{1}{T}\sum_t^T S_tF_{tn}.
\end{equation}
Obviously, $G^{(2)}_{n}$ is also a stochastic variable. Now, let us calculate its mean:
\begin{align}\label{eq:EG2}
E(G^{(2)}_{n})&=\frac{1}{T}\sum_t^T E(S_tF_{tn})\nonumber\\
		      &=\frac{1}{T}\sum_t^T E[(\widetilde{S}_{t}+\gamma d_nI_{tn})F_{tn}]\nonumber\\
		      &=\frac{1}{T}\sum_t^T [E(\widetilde{S}_{t})E(F_{tn})+E(\gamma d_nI_{tn}F_{tn})].
\end{align}
For the first term of Eq.~(\ref{eq:EG2}), it can be written as
\begin{align}\label{eq:firstterm}
E(\widetilde{S}_{t})E(F_{tn})&=E(\gamma \sum_{m\ne n}^M d_mI_{tm})E(F_{tn})\nonumber\\
		&=\gamma\sum_{m\ne n}^M d_mE(\mathcal{I})E(\mathcal{F})\nonumber\\
		&=\gamma\sum_m^M d_mE(\mathcal{I})E(\mathcal{F})-\gamma d_nE(\mathcal{I})E(\mathcal{F}).
\end{align}
For the second term of Eq.~(\ref{eq:EG2}), it can be given as
\begin{align}\label{eq:BE}
E(\gamma d_mI_{tm}F_{tn})&=\gamma d_mE(I_{tm}F_{tn})\nonumber\\
		       &=\begin{cases}
			     \gamma d_nE(\mathcal{IF})& m=n\\
			     \gamma d_mE(\mathcal{I})E(\mathcal{F})& m\ne n,
		         \end{cases}
\end{align}
which we call the independence discrimination separated formula. Substituting Eqs.~(\ref{eq:firstterm}) and (\ref{eq:BE}) into Eq.~(\ref{eq:EG2}) gives
\begin{align}\label{eq:EG2final}
E(G^{(2)}_{n})=&\frac{1}{T}\sum_t^T[\gamma\sum_m^M d_mE(\mathcal{I})E(\mathcal{F})-\gamma d_nE(\mathcal{I})E(\mathcal{F})\nonumber\\
               &+\gamma d_nE(\mathcal{IF})]\nonumber\\
		      =&\gamma\sum_m^M d_mE(\mathcal{I})E(\mathcal{F})\nonumber\\
               &+\gamma[E(\mathcal{IF})-E(\mathcal{I})E(\mathcal{F})]d_n\nonumber\\
		      =&C_2+C_1d_n,
\end{align}
where both $C_1$ and $C_2$ are constants:
\begin{align}
C_1&=\gamma[E(\mathcal{IF})-E(\mathcal{I})E(\mathcal{F})],\\
C_2&=\gamma\sum_m^M d_mE(\mathcal{I})E(\mathcal{F}).
\end{align}

Therefore, the mean of $G^{(2)}_{n}$ is a linear transformation of the original object's gray value $d_n$, while the transformation coefficients $C_1$ and $C_2$ are independent of this gray value $d_n$. This explains why the object image can be restored. We can also see that for different functional forms $\mathcal{F}$ of $\mathcal{I}$, it only affects the coefficients $C_1$ and $C_2$, and the reconstruction transformation is still linear, thus the image can still be recovered. This also explains why power functions, exponential functions, logarithmic functions, etc. can be used to reconstruct the ghost images.

Following above calculation idea, we further deduce the mean of the background-removal correlation function $\Delta G^{(2)}$. For the $n$th pixel, $\Delta G^{(2)}_n$ can be written as
\begin{align}
\Delta G^{(2)}_n&=\langle S\cdot F_n\rangle-\langle S\rangle\langle F_n\rangle\nonumber\\
		&=G^{(2)}_n-\frac{1}{T^2}\sum_t^T S_t\sum_t^T F_{tn}\nonumber\\
		&=G^{(2)}_n-\frac{1}{T^2}(\sum_t^T S_tF_{tn}+\sum_t^T\sum_{t'\ne t}^T S_tF_{t'n})\nonumber\\
		&=G^{(2)}_n-\frac{1}{T}G^{(2)}_n-\frac{1}{T^2}\sum_t^T\sum_{t'\ne t}^T S_tF_{t'n}.
\end{align}
Then, we calculate its mean:
\begin{equation}\label{eq:DeltaG2n}
E(\Delta G^{(2)}_n)=(1-\frac{1}{T})E(G^{(2)}_n)-\frac{1}{T^2}E(\sum_t^T\sum_{t'\ne t}^T S_tF_{t'n}).
\end{equation}
Since the first term has already been given by Eq.~(\ref{eq:EG2final}), we will directly calculate the second term:
\begin{align}\label{eq:Deltasecondterm}
E(\sum_t^T\sum_{t'\ne t}^T S_tF_{t'n})&=\sum_t^T\sum_{t'\ne t}^T E(S_tF_{t'n})\nonumber\\
&=\sum_t^T\sum_{t'\ne t}^T E(S_t)E(F_{t'n})\nonumber\\
&=\sum_t^T\sum_{t'\ne t}^T(\gamma\sum_m^M d_mE(I_{tm}))E(F_{t'n})\nonumber\\
&=T(T-1)\gamma\sum_m^M d_m E(\mathcal{I})E(\mathcal{F}).
\end{align}
Substitute Eqs.~(\ref{eq:EG2final}) and (\ref{eq:Deltasecondterm}) into Eq.~(\ref{eq:DeltaG2n}), then we will have
\begin{align}\label{eq:Deltafinal}
E(\Delta G^{(2)}_n)=&\left(1-\frac{1}{T}\right)\{\gamma\sum_m^M d_mE(\mathcal{I})E(\mathcal{F})\nonumber\\
                    &+\gamma[E(\mathcal{IF})-E(\mathcal{I})E(\mathcal{F})]d_n\}\nonumber\\
                    &-\left(1-\frac{1}{T}\right)\gamma\sum_m^M d_m E(\mathcal{I})E(\mathcal{F})\nonumber\\
                   =&\left(1-\frac{1}{T}\right)\gamma[E(\mathcal{IF})-E(\mathcal{I})E(\mathcal{F})]d_n\nonumber\\
             \approx&\gamma[E(\mathcal{IF})-E(\mathcal{I})E(\mathcal{F})]d_n\nonumber\\
                   =&C_1d_n.
\end{align}
Obviously, the mean of $\Delta G^{(2)}_{n}$ is also a linear transformation of the original gray value $d_n$, while the transformation coefficient $C_1$ is also independent of this gray value $d_n$. Therefore, the object information can be retrieved by $\Delta G^{(2)}_{n}$. For different functions $\mathcal{F}=f(\mathcal{I})$, this linear transformation relationship still exists.

In the above, we have already analyzed the mean of $\Delta G^{(2)}_{n}$, but this analysis is not enough. As we know, there exists some certain noise fluctuations in the reconstructed result of $\Delta G^{(2)}_{n}$, so only using the average value to describe the reconstruction performance of $\Delta G^{(2)}_{n}$ is one-sided. Next, we will discuss such fluctuations in detail. As we know, $\Delta G^{(2)}_n$ can be derived from second-order correlation function $G^{(2)}_n$:
\begin{align}\label{eq:Deltaequivalence}
\Delta G^{(2)}_n=&\langle SF_n\rangle-\langle S\rangle\langle F_n\rangle\nonumber\\
\Leftrightarrow&\langle(S-\langle S\rangle)(F_n-\langle F_n\rangle)\rangle.
\end{align}
When the number of reference patterns is large enough, there are $\langle S\rangle\approx E(S)$ and $\langle F_n\rangle\approx E(F_n)$. Thus, Eq.~(\ref{eq:Deltaequivalence}) can be approximated as
\begin{align}
\Delta G^{(2)}_n\approx&\langle(S-E(S))(F_n-E(F_n))\rangle\nonumber\\
	                  =&\langle(S-E(S))(F_n-E(\mathcal{F}))\rangle.
\end{align}

According to the well-known ``central limit theorem" \cite{Laplace1812,Lyapunov1954} in probability theory: when the number of reference patterns $T$ is large enough, the result of $\Delta G^{(2)}_n$ for an original gray value $d_n$ will obey a Gaussian distribution with a mean $\mu_n=E[(S-E(S))(F_n-E(\mathcal{F}))]$ and a variance $\sigma_n^2=\frac{1}{T}D\{[S-E(S)][F_n-E(\mathcal{F})]\}$. This reveals an important feature of $\Delta G^{(2)}$: for a certain gray-scale value of the original object, the calculated $\Delta G^{(2)}_n$ value is not a certain number, but follows a Gaussian distribution. That is, the reconstructed $\Delta G^{(2)}_n$ values for different object gray-scale values will fluctuate around different means, obeying different Gaussian distributions, and their differences can be resolved from the recovered image. The feature that ``one original gray-scale level corresponds to one Gaussian curve, which can be calculated from the reconstructed values in the corresponding pixel positions" is crucial for explaining why $\Delta G^{(2)}_n$ can reconstruct a ghost image.

\section{\label{sec:level3}Numerical simulation}
To verify that the recovered values of deformation function form $\Delta G^{(2)}_n$ with respect to each original gray value can obey a certain theoretical distribution, we use the patterns that obey a given distribution $\mathcal{I}$ for measurement and applied its functional forms $\mathcal{F}=f(\mathcal{I})$ for reconstruction. Since we cannot go through all functions for verification, only some common functions such as $\mathcal{F}=\mathcal{I}$, $\mathcal{F}=\mathcal{I}^3$, $\mathcal{F}=exp(\mathcal{I})$ and $\mathcal{F}=ln(\mathcal{I})$ are used here. After reconstructing the image, we calculate the practical probability density distributions and the Gaussian theoretical curves (obtained from theoretical mean and variance) of recovered pixel values falling in each pixel region of the same original gray value $d_n$, to see whether the reconstructed data is consistent with the theoretical Gaussian curve. In the previous section, we have pointed out that the mean and variance of theoretical Gaussian curve by using functional form $\Delta G^{(2)}_n$ to calculate the pixels of original gray value $d_n$ are $\mu_n=E[(S-E(S))(F_n-E(\mathcal{F}))]$ and $\sigma_n^2=\frac{1}{T}D\{[S-E(S)][F_n-E(\mathcal{F})]\}$. Now, we need to express both the mean and variance in terms of those of $\mathcal{I}$ and $\mathcal{F}$.

For the mean $\mu_n=E[(S-E(S))(F_n-E(\mathcal{F}))]$, there is
\begin{align}
 &E[(S-E(S))(F_n-E(\mathcal{F}))]\nonumber\\
=&E(SF_n)-E(S)E(\mathcal{F})\nonumber\\
=&E[(\widetilde{S}+\gamma d_nI_n)F_n]-E(\widetilde{S}+\gamma d_nI_n)E(\mathcal{F})\nonumber\\
=&E(\widetilde{S})E(\mathcal{F})+\gamma d_nE(\mathcal{IF})\nonumber\\
 &-(E(\widetilde{S})E(\mathcal{F})+\gamma d_nE(\mathcal{I})E(\mathcal{F}))\nonumber\\
=&\gamma[E(\mathcal{IF})-E(\mathcal{I})E(\mathcal{F})]d_n.
\end{align}

For the variance $\sigma_n^2=\frac{1}{T}D\{[S-E(S)][F_n-E(\mathcal{F})]\}$, we have
\begin{align}
 &D\{[S-E(S)][F_n-E(\mathcal{F})]\}\nonumber\\
=&D[SF_n-E(S)F_n-E(\mathcal{F})S]\nonumber\\
=&D(SF_n)+D[E(S)F_n]+D[E(\mathcal{F})S]\nonumber\\
 &-2Cov[SF_n,E(S)F_n]-2Cov[SF_n,E(\mathcal{F})S]\nonumber\\
 &+2Cov[E(S)F_n,E(\mathcal{F})S]\nonumber\\
=&D(SF_n)+E(S)^2D(\mathcal{F})+E(\mathcal{F})^2D(S)\nonumber\\
 &-2E(S)[E(SF_n^2)-E(SF_n)E(\mathcal{F})]\nonumber\\
 &-2E(\mathcal{F})[E(S^2F_n)-E(SF_n)E(S)]\nonumber\\
 &+2E(S)E(\mathcal{F})[E(SF_n)-E(S)E(\mathcal{F})]\nonumber\\
=&E(S)[6E(SF_n)E(\mathcal{F})-2E(SF_n^2)]\nonumber\\
 &+E(S)^2[D(\mathcal{F})-2E(\mathcal{F})^2]\nonumber\\
 &+D(S)E(\mathcal{F})^2-2E(S^2F_n)E(\mathcal{F})+D(SF_n).
\end{align}
Obviously, there are still many terms in the formula of variance that have not been expressed by the mean and variance of $\mathcal{I}$ and $\mathcal{F}$. To this end, we will calculate analyze each item in this variance formula. The unknown terms $E(S)$, $E(SF_n)$ and $E(SF_n^2)$ in the first term $E(S)[6E(SF_n)E(\mathcal{F})-2E(SF_n^2)]$ of the variance formula can be calculated as follows
\begin{align}
E(S)=&E(\widetilde{S}+\gamma d_nI_n)=E(\widetilde{S})+\gamma d_nE(\mathcal{I}),\label{eq:ES}\\
E(SF_n)=&E[(\widetilde{S}+\gamma d_nI_n)F_n]\nonumber\\
       =&E(\widetilde{S})E(\mathcal{F})+\gamma d_nE(\mathcal{IF}),\label{eq:SFn}\\
E(SF_n^2)=&E[(\widetilde{S}+\gamma d_nI_n)F_n^2]\nonumber\\
         =&E(\widetilde{S})E(\mathcal{F}^2)+\gamma d_nE(\mathcal{I}\mathcal{F}^2),
\end{align}
where
\begin{equation}\label{eq:EtildeS}
E(\widetilde{S})=E(\gamma\sum_{m\ne n}^M d_mI_m)=\gamma\sum_{m\ne n}^M d_mE(\mathcal{I}).
\end{equation}
Since the second term $E(S)^2[D(\mathcal{F})-2E(\mathcal{F})^2]$ of the variance formula only has one unknown term $E(S)$, which has been provided by Eq.~(\ref{eq:ES}). The unknown term $D(S)$ in the third term $D(S)E(\mathcal{F})^2$ of the variance formula can be deduced as
\begin{equation}
D(S)=D(\widetilde{S}+\gamma d_nI_n)=D(\widetilde{S})+\gamma^2 d_n^2D(\mathcal{I}),
\end{equation}
where
\begin{equation}\label{eq:DtildeS}
D(\widetilde{S})=D(\gamma\sum_{m\ne n}^M d_mI_m)=\gamma^2\sum_{m\ne n}^M d_m^2D(\mathcal{I}).
\end{equation}
Then, we compute the unknown term $E(S^2F_n)$ in the fourth term $-2E(S^2F_n)E(\mathcal{F})$:
\begin{align}
E(S^2F_n)=&E[(\widetilde{S}+\gamma d_nI_n)^2F_n]\nonumber\\
	     =&E[(\widetilde{S}^2+2\widetilde{S}\gamma d_nI_n+\gamma^2 d_n^2I_n^2)F_n]\nonumber\\
	     =&E(\widetilde{S}^2F_n)+2\gamma d_nE(\widetilde{S}I_nF_n)+\gamma^2 d_n^2E(I_n^2F_n)\nonumber\\
	     =&E(\widetilde{S}^2)E(\mathcal{F})+2\gamma d_nE(\widetilde{S})E(\mathcal{IF})\nonumber\\
          &+\gamma^2 d_n^2E(\mathcal{I}^2\mathcal{F}).
\end{align}
where $E(\widetilde{S})$ has been provided by Eq.~(\ref{eq:EtildeS}), and $E(\widetilde{S})^2$ can be calculated by using Eqs.~(\ref{eq:EtildeS}) and (\ref{eq:DtildeS}):
\begin{align}\label{eq:EtildeS2}
E(\widetilde{S}^2)=&E(\widetilde{S})^2+D(\widetilde{S})\nonumber\\
=&\left[\gamma\sum_{m\ne n}^M d_mE(\mathcal{I})\right]^2+\gamma^2\sum_{m\ne n}^M d_m^2D(\mathcal{I}).
\end{align}
Now, let us calculate the fifth term $D(SF_n)$:
\begin{equation}
D(SF_n)=E(S^2F_n^2)-E(SF_n)^2.
\end{equation}
where $E(SF_n)$ has been provided by Eq.~(\ref{eq:SFn}), and $E(S^2F_n^2)$ can be given by
\begin{align}\label{eq:ES2Fn2}
E(S^2F_n^2)=&E[(\widetilde{S}+\gamma d_nI_n)^2F_n^2]\nonumber\\
	       =&E[(\widetilde{S}^2+2\widetilde{S}\gamma d_nI_n+\gamma^2 d_n^2I_n^2)F_n^2]\nonumber\\
	       =&E(\widetilde{S}^2F_n^2)+2\gamma d_nE(\widetilde{S}I_nF_n^2)+\gamma^2 d_n^2E(I_n^2F_n^2)\nonumber\\
	       =&E(\widetilde{S}^2)E(\mathcal{F}^2)+2\gamma d_n E(\widetilde{S})E(\mathcal{I}\mathcal{F}^2)\nonumber\\
            &+\gamma^2d_n^2E(\mathcal{I}^2\mathcal{F}^2).
\end{align}
In above formula, $E(\widetilde{S})$ and $E(\widetilde{S}^2)$ has been given by Eqs.~(\ref{eq:EtildeS}) and (\ref{eq:EtildeS2}), respectively. So far, we have expressed each term in the variance formula in terms of the mean and variance of $\mathcal{I}$ and $\mathcal{F}$. Now, all theoretical calculations for the mean $\mu_n=E[(S-E(S))(F_n-E(\mathcal{F}))]$ and the variance $\sigma_n^2=\frac{1}{T}D\{[S-E(S)][F_n-E(\mathcal{F})]\}$ are completed, all described by the mean and variance of $\mathcal{I}$ and $\mathcal{F}$. Based on the mean and variance, we can easily acquire the Gaussian theoretical curves of recovered pixel values for each original gray value $d_n$.

\begin{figure}[htbp]
\centering
\includegraphics[width=0.9\linewidth]{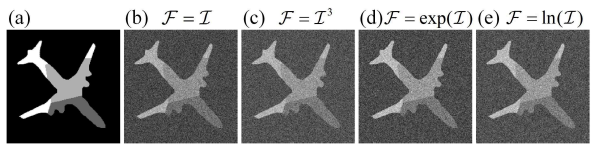}
\caption{\label{fig:simulation}Simulation results. (a) is the original airplane image with four different gray-scale values. (b)--(e) are the reconstructed images using different functions $\mathcal{F}=\mathcal{I}$, $\mathcal{F}=\mathcal{I}^3$, $\mathcal{F}=exp(\mathcal{I})$ and $\mathcal{F}=ln(\mathcal{I})$.}
\end{figure}

\begin{figure*}[htbp]
\centering
\includegraphics[width=0.9\linewidth]{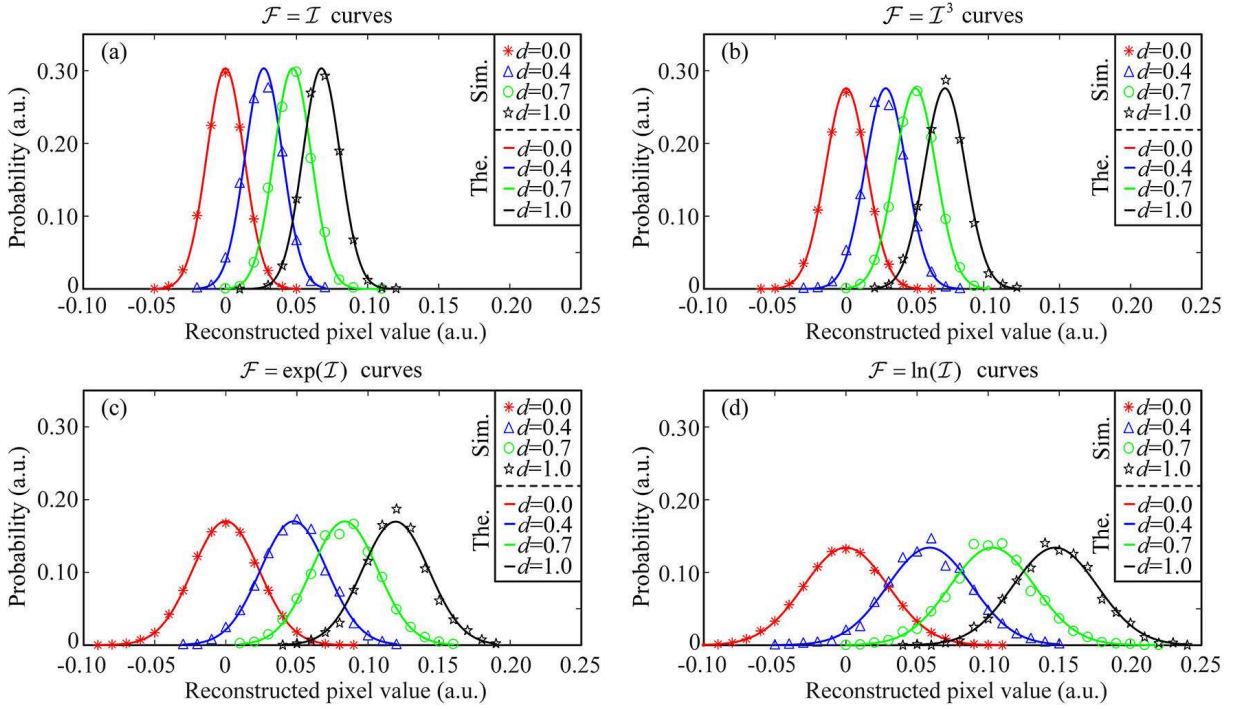}
\caption{\label{fig:PDF}Probability versus the recovered pixel values falling in the pixel region where the original gray value $d=$0, 0.4, 0.7 and 1.0, compared with the Gaussian theoretical curves. (a)--(d) are the cases of using $\mathcal{F}=\mathcal{I}$, $\mathcal{F}=\mathcal{I}^3$, $\mathcal{F}=exp(\mathcal{I})$ and $\mathcal{F}=ln(\mathcal{I})$ in $\Delta G^{(2)}_n$, respectively. The abbreviations ``Sim." and ``The." separately stand for the simulation data and the theoretical curve.}
\end{figure*}

The object image used for numerical simulation is an airplane graph of $200\times200$ pixels created by us, containing four gray-scale values 0, 0.4, 0.7 and 1.0, which separately account for 81.975\%, 5.265\%, 8.055\%, and 4.705\% of the whole pixels, as shown in Fig.~\ref{fig:simulation}(a). We generated 100000 reference patterns which obeyed an identical uniform distribution, with the values ranging from 0.1 to 1. Accordingly, 100000 simulated measurements could be acquired. Here, we used $\mathcal{F}=\mathcal{I}$, $\mathcal{F}=\mathcal{I}^3$, $\mathcal{F}=exp(\mathcal{I})$ and $\mathcal{F}=ln(\mathcal{I})$ in $\Delta G^{(2)}_n$, and the corresponding reconstructed results were given in Fig.~\ref{fig:simulation}(b)--\ref{fig:simulation}(e). We counted the value probability of reconstructed pixels in the region where the pixels of the same original gray value were located, with comparisons to their Gaussian theoretical curves. The Gaussian theoretical curves and calculated probability statistics for different functions were presented in Figs.~\ref{fig:PDF}(a)--\ref{fig:PDF}(d), where the ordinate represented the occurrence probability of these reconstructed values. It could be clearly seen that the probability of recovered pixel values were highly consistent with their Gaussian theoretical distributions.

\section{\label{sec:level4}Experiment and results}
\begin{figure}[htbp]
\centering
\includegraphics[width=0.95\linewidth]{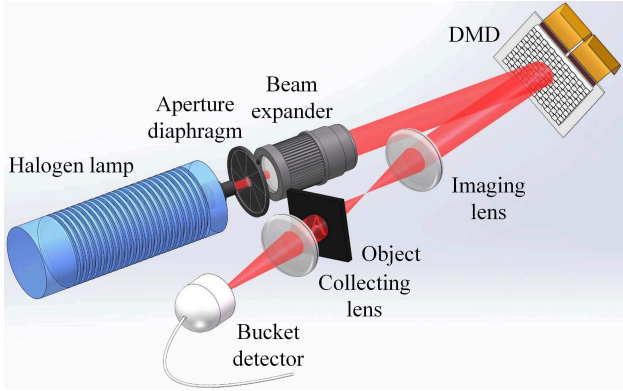}
\caption{\label{fig:setup}Experimental apparatus for computational GI with a DMD. The computational illumination of thermal light generated by the light intensity modulation of the DMD was projected onto an object. The total transmission light intensities were collected by a bucket detector.}
\end{figure}

In experiment, we applied a computational GI scheme, in which a DMD was used as a SLM, as shown in Fig.~\ref{fig:setup}. The thermal light from a halogen lamp passed through an aperture diaphragm and was collimated by a beam expander. Then, the light beam illuminated the DMD, with the light intensity being modulated by the preset patterns of the latter. The used DMD consisted of 1024 $\times$ 768 micromirrors, each of which was of size $13.68\times13.68\ \mu\textrm{m}^2$ and could be switched to either +12$^\circ$ and -12$^\circ$ based on the pixel value 1 or 0 on the modulated patterns. Thereby, we let the illumination light be incident to the DMD working plane at an angle of 24$^\circ$ from its normal, so that the light corresponding to the bright pixel 1 could be emitted along the normal direction and was projected onto a black-and-white film printed with the letter ``A" (as an object). The transmission light then converged to a bucket (single-pixel) detector through a collecting lens. Here, we used uniformly distributed 0-1 modulation patterns (obeying an identical distribution $\mathcal{I}$) for measurements, each pattern occupied the central $160\times160$ pixels of the DMD. Thus, the probability of occurring 0 and 1 on the modulated patterns is the same, both of 0.5. The functions chosen for calculations are: $\mathcal{F}=\mathcal{I}$, $\mathcal{F}=\mathcal{I}^3$, and $\mathcal{F}=exp(\mathcal{I})$. The reason why $\mathcal{F}=ln(\mathcal{I})$ was not used here was that $ln(0)$ did not exist when the reference pattern pixels could take the values of 0.
\begin{figure}[htbp]
\centering
\includegraphics[width=0.9\linewidth]{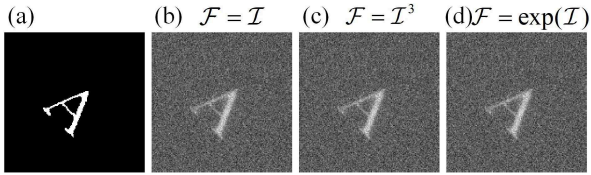}
\caption{\label{fig:expresults}Experimental results. (a) is a binarized image taken by a camera. (b)--(d) are the recovered images using $\mathcal{F}=\mathcal{I}$, $\mathcal{F}=\mathcal{I}^3$, and $\mathcal{F}=exp(\mathcal{I})$, respectively.}
\end{figure}
\begin{figure*}[htbp]
\centering
\includegraphics[width=0.9\linewidth]{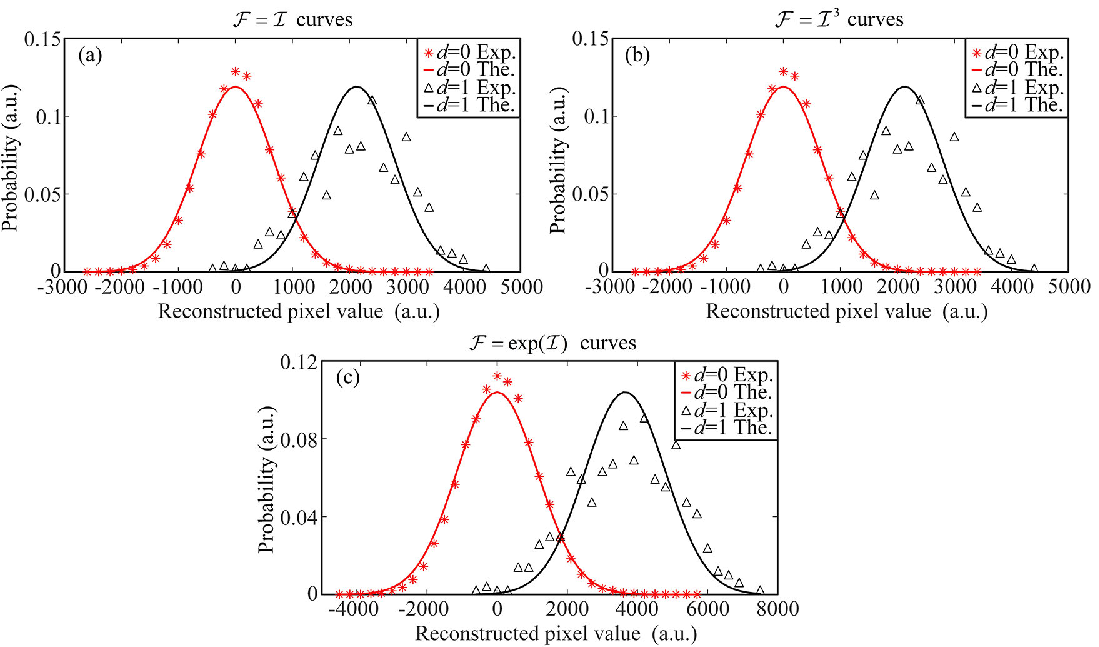}
\caption{\label{fig:expPDF}Probability as a function of recovered pixel values that locate in the pixel region where the original gray value equals 0 or 1, compared with their Gaussian theoretical curves. The calculations use (a) $\mathcal{F}=\mathcal{I}$, (b) $\mathcal{F}=\mathcal{I}^3$ and (c) $\mathcal{F}=exp(\mathcal{I})$ in the function $\Delta G^{(2)}_n$. The abbreviation ``Exp." stands for the experimental data and ``The." is short for the theoretical curve.}
\end{figure*}

The experimental results of $\Delta G^{(2)}_n$ using 11940 measurements were given in Figs.~\ref{fig:expresults}(b)--\ref{fig:expresults}(d). Since the measurement noise was inevitable, its influence should be considered. Given this, the bucket value can be written as $S'=S+e=\widetilde{S}+\gamma d_nI_n+e=(\widetilde{S}+e)+\gamma d_nI_n$, where $e$ denotes the noise. Here, we only need to replace $\widetilde{S}$ in Eqs.~(\ref{eq:ES})--(\ref{eq:ES2Fn2}) with $\widetilde{S}+e$ for acquiring the noisy formulas. We will have the mean $E(\widetilde{S}+e)=E(\widetilde{S})+E(e)=\gamma\sum_{m\ne n}^M d_mE(\mathcal{I})+E(e)$ and the variance $D(\widetilde{S}+e)=D(\widetilde{S})+E(e)=\gamma^2\sum_{m\ne n}^M d_m^2D(\mathcal{I})+D(e)$. By estimation, the average value of the measurement noise was $E(e)=2.0985\times10^{6}$, and its variance was $D(e)=1.2260\times10^{10}$. Then, we calculated the probability of reconstructed pixel values that located in the region in which the original gray-scale value is 0 or 1, and made an comparison with the Gaussian theoretical curves, as shown in Fig.~\ref{fig:expPDF}. From the charts, it could be seen clearly that the experimental statistical data agreed well with the Gaussian theoretical curves.

\section{\label{sec:level5}Extension: to explain other correlation functions}
By using the same idea, it can also be proven that the mean results of other traditional intensity correlation functions of GI are also the linear transformations of the original object's gray values. We can take $g^{(2)}_n=\frac{\langle S\cdot F_n\rangle}{\langle S\rangle\langle F_n\rangle}$ and $\textrm{DGI}_n=\langle S\cdot F_n\rangle-\frac{\langle S\rangle}{\langle S_R\rangle}\langle S_R\cdot F_n\rangle$ ($S_R$ is defined as $S_R=\sum_m^M I_m$) \cite{Ferri2010} for example.

For $g^{(2)}_n$, there is
\begin{align}
g^{(2)}_n=&\frac{\langle S\cdot F_n\rangle}{\langle S\rangle\langle F_n\rangle}\nonumber\\
   \approx&\frac{\langle S\cdot F_n\rangle}{E(S)E(F_n)}\nonumber\\
		 =&\frac{G^{(2)}_n}{\gamma\sum_m^M d_mE(\mathcal{I})E(\mathcal{F})}.
\end{align}
Using $E(G^{(2)}_n)$ result, we can obtain the mean of $g^{(2)}_n$:
\begin{align}
 &E(g^{(2)}_n)\nonumber\\
=&\frac{\gamma\sum_m^M d_mE(\mathcal{I})E(\mathcal{F})+\gamma[E(\mathcal{IF})-E(\mathcal{I})E(\mathcal{F})]d_n}{\gamma\sum_m^M d_mE(\mathcal{I})E(\mathcal{F})}\nonumber\\
=&1+\frac{E(\mathcal{IF})-E(\mathcal{I})E(\mathcal{F})}{\sum_m^M d_mE(\mathcal{I})E(\mathcal{F})}d_n\nonumber\\
=&1+C_3d_n,
\end{align}
where $C_3=\frac{E(\mathcal{IF})-E(\mathcal{I})E(\mathcal{F})}{\sum_m^M d_mE(\mathcal{I})E(\mathcal{F})}$.

For $\textrm{DGI}_n$, we have
\begin{align}
\textrm{DGI}_n=&\langle S\cdot F_n\rangle-\frac{\langle S\rangle}{\langle S_R\rangle}\langle S_R\cdot F_n\rangle\nonumber\\
        \approx&\langle S\cdot F_n\rangle-\frac{E(S)}{E(S_R)}\langle S_R\cdot F_n\rangle\nonumber\\
	          =&G^{(2)}_n-\frac{\gamma\sum_m^M d_mE(\mathcal{I})}{ME(\mathcal{I})}\langle S_R\cdot F_n\rangle,
\end{align}
where only the mean of the term $\langle S_R\cdot F_n\rangle$ is unknown. We can deduce this mean as
\begin{align}
E(\langle S_R\cdot F_n\rangle)=&E(\sum_m^M I_mF_n)\nonumber\\
                              =&E[(\sum_{m\ne n}^M I_m+I_n)F_n]\nonumber\\
		                      =&(M-1)E(\mathcal{I})E(\mathcal{F})+E(\mathcal{IF})\nonumber\\
                              =&ME(\mathcal{I})E(\mathcal{F})+E(\mathcal{IF})-E(\mathcal{I})E(\mathcal{F}).
\end{align}
Thus, the mean of $\textrm{DGI}_n$ can be written as
\begin{align}
 &E(\textrm{DGI}_n)\nonumber\\
=&E(G^{(2)}_n)-\frac{\gamma\sum_m^M d_mE(\mathcal{I})}{ME(\mathcal{I})}E(\langle S_R\cdot F_n\rangle)\nonumber\\
=&\gamma\sum_m^M d_mE(\mathcal{I})E(\mathcal{F})+\gamma[E(\mathcal{IF})-E(\mathcal{I})E(\mathcal{F})]d_n\nonumber\\
 &-\frac{\gamma\sum_m^M d_m}{M}[ME(\mathcal{I})E(\mathcal{F})+E(\mathcal{IF})-E(\mathcal{I})E(\mathcal{F})]\nonumber\\
=&\gamma[E(\mathcal{IF})-E(\mathcal{I})E(\mathcal{F})]\left(d_n-\frac{\sum_m^M d_m}{M}\right)\nonumber\\
=&C_1(d_n-C_4),
\end{align}
where $C_1=\gamma[E(\mathcal{IF})-E(\mathcal{I})E(\mathcal{F})]$ and $C_4=\frac{\sum_m^M d_m}{M}$.

Obviously, no matter which kind of function $\mathcal{F}=f(\mathcal{I})$ is used, the means of both $g^{(2)}_n$ and $\textrm{DGI}_n$ are the linear transformations of original gray values. Therefore, the target gray information is retained, and the object image can be recovered.

\section{\label{sec:level6}Conclusion}
In summary, we have proposed and demonstrated a unified theoretical model that assumes the reference patterns of arbitrary identical distribution $\mathcal{I}$ and the objects of gray-scale, to reveal the imaging mechanism of correlation functions. Take $\Delta G_2$ for example, no matter which kind of functional forms $\mathcal{F}=f(\mathcal{I})$ is used for calculation, the probability of recovered pixel values that locate in the pixel region of the same original gray value will present a Gaussian distribution. The means of these Gaussian distributions for different pixel regions have a linear relationship with their original gray values, which explains why intensity correlation can retrieve the object information. Each Gaussian distribution has a variance, which indicates the fluctuation of the reconstructed pixel values falling in the same pixel region, and accounts for the facts that why the visibility of GI is not very high, generally accompanied by a lot of noise. As a proof and promotion of concept, other two classic correlation functions $g^{(2)}_n$ and $\textrm{DGI}_n$ were discussed to further verify the universality of this theory. According to our strict theoretical proofs, the essential reason why a classical correlation function can reconstruct the object image is that the reconstruction mean in a specified pixel region (corresponding to the same original gray value) has a linear relationship with this original gray value. Thereby, this work gives a statistical perspective to the GI theory.

\begin{acknowledgments}
This work was supported by the National Natural Science Foundation of China (Grant No. 61801022), the Natural Science Foundation of Beijing Municipality (Grant No. 4184098), the National Key Research and Development Program of China (Grant No. 2016YFE0131500), the Civil Space Project of China (Grant No. D040301), the International Science and Technology Cooperation Special Project of Beijing Institute of Technology (Grant No. GZ2018185101).
\end{acknowledgments}

\nocite{*}


\end{document}